\documentstyle[12pt,epsfig]{article}

\newcommand{\be}{\begin{equation}}
\newcommand{\ee}{\end{equation}}
\newcommand{\Dlt}{\Delta}
\newcommand{\dlt}{\delta}
\newcommand{\prt}{\partial}
\newcommand{\br}{{\bf r}}
\newcommand{\bk}{{\bf k}}
\newcommand{\bt}{\beta}
\newcommand{\vp}{\varphi}
\newcommand{\ep}{\varepsilon}

\newcommand{\ra}{\rightarrow}
\newcommand{\sgm}{\sigma}
\newcommand{\gm}{\gamma}
\newcommand{\om}{\omega}
\newcommand{\Om}{\Omega}

\newcommand{\dgr}{\dagger}
\newcommand{\lbd}{\lambda}
\newcommand{\Lbd}{\Lambda}
\newcommand{\cF}{{\cal F}}

\begin{document}

\begin{center}

{\Large{\bf Bose-Einstein-condensed gases with arbitrary strong interactions} \\ [5mm]

V.I. Yukalov$^1$ and E.P. Yukalova$^2$} \\ [3mm]

{\it $^1$Bogolubov Laboratory of Theoretical Physics, \\
Joint Institute for Nuclear Research, Dubna 141980, Russia \\ [3mm]
$^2$Department of Computational Physics,
Laboratory of Information Technologies, \\
Joint Institute for Nuclear Research, Dubna 141980, Russia}

\end{center}

\begin{abstract}

Bose-condensed gases are considered with an effective interaction
strength varying in the whole range of the values between zero and infinity.
The consideration is based on the usage of a representative statistical
ensemble for Bose systems with broken global gauge symmetry. Practical
calculations are illustrated for a uniform Bose gas at zero temperature,
employing a self-consistent mean-field theory, which is both conserving and
gapless.

\end{abstract}

\vskip 1cm


\newpage

\section{Introduction}

The properties of systems with Bose-Einstein condensate are currently a topic
of great interest, both experimentally and theoretically (see review articles
[1--7]). One usually considers weakly interacting Bose gases, whose theory was
pioneered by Bogolubov [8,9]. Binary atomic interactions in such gases can be
modelled by contact potentials expressed through effective scattering lengths.
But the latter can also be made rather large by means of the Feshbach resonance
technique, so that effective atomic interactions could become quite strong
[7,10,11]. Extension of the Bogolubov theory to Bose systems with strong
interactions confronts the well known problem of conserving versus gapless
approximations, as was formulated by Hohenberg and Martin [12]. This dilemma
has recently been discussed in detail in the review paper by Andersen [5].

The Hohenberg-Martin dilemma of conserving versus gapless theories can be
resolved by employing representative statistical ensembles [13]. Using such
an ensemble for Bose systems with broken global gauge symmetry makes it
straightforward to get a self-consistent theory, both conserving as well as
gapless in any given approximation. In particular, the Hartree-Fock-Bogolubov
(HFB) approximation, which is by construction conserving, can also be made
gapless [14].

In the present paper, we consider an equilibrium Bose system with Bose-Einstein
condensate. The main new results are twofold. First, we give a general
mathematical foundation for the construction of the grand Hamiltonian for
an arbitrary equilibrium system with broken gauge symmetry. The derivation
of the grand Hamiltonian and the corresponding equations of motion are valid
for any Bose system, whether uniform or nonuniform. Second, in the frame of
a self-consistent mean-field theory for a uniform Bose gas, we study, both
analytically and numerically, the zero-temperature characteristics as functions
of the gas parameter, varying the latter from zero to infinity. Specifically,
the condensate fraction, sound velocity, normal and anomalous averages, and
the ground-state energy as functions of the gas parameter are investigated.
The results are in good agreement with available computer Monter Carlo
simulations.

We use the system of units, where $\hbar\equiv 1$ and $k_B\equiv 1$.

\section{Representative Ensemble for Bose-Condensed Systems}

The description of a spinless Bose system at temperature $T>T_c$ above the
condensation temperature $T_c$ can be done in terms of the field operators
$\psi(\br,t)$ and $\psi^\dgr(\br,t)$ dependeing on the spatial vector $\br$
and time $t$. The operators from the algebra of observables and other
physical operators are defined in the Fock space $\cF(\psi)$ generated by the
field operator $\psi^\dgr$. The related mathematical details of constructing
the Fock space $\cF(\psi)$ can be found in books [15,16]. Under the total
number of particles $N=<\hat N>$, being the average of the number-of-particle
operator $\hat N$, the grand Hamiltonian has the standard form
$$
H[\psi] = \hat H [\psi] - \mu\hat N \qquad (T > T_c) \; ,
$$
where $\hat H[\psi]$ is the Hamiltonian energy, which is invariant under the
global gauge transformations from the $U(1)$ symmetry group.

At temperatures $T<T_c$, the global gauge symmetry becomes broken. This is
achieved by means of the Bogolubov shift [17,18] for the field operators
\be
\label{1}
\psi(\br,t) \; \longrightarrow \; \hat\psi(\br,t)\equiv \eta(\br,t)
+ \psi_1(\br,t) \; ,
\ee
in which $\eta(\br,t)$ is the condensate wave function and $\psi_1(\br,t)$
is the field operator of uncondensed atoms, enjoying the same Bose
commutation relations as $\psi$. The condensate wave function $\eta(\br,t)$
is the system order parameter. Now all operators of physical quantities are
defined on the Fock space $\cF(\psi_1)$ generated by the field operator
$\psi_1^\dgr$. It is important to emphasize that the Fock spaces $\cF(\psi)$
and $\cF(\psi_1)$ are mutually orthogonal [13,19].

Thus, below $T_c$, instead of one operator variable $\psi$, there appear two
variables, $\eta$ and $\psi_1$. These are linearly independent, being orthogonal
to each other,
\be
\label{2}
\int \eta^*(\br,t) \psi_1(\br,t) \; d\br = 0 \; .
\ee
For two linearly independent variables, there are two normalization conditions.
One is the normalization of the condensate function to the number of
condensed atoms
\be
\label{3}
N_0 = \int |\eta(\br,t)|^2 \; d\br \; .
\ee
And another normalization condition is for the operator
\be
\label{4}
\hat N_1 \equiv \int \psi_1^\dgr(\br,t) \psi_1(\br,t)\; d\br \; ,
\ee
whose average yields the number of uncondensed atoms
\be
\label{5}
N_1 \;  = \; <\hat N_1> \; .
\ee
The normalization condition (3) can be represented in the same form of the
statistical average (5) by using the operator
$$
\hat N_0 \equiv N_0 \hat 1_\cF \; ,
$$
where $\hat 1_\cF$ is the unity operator in $\cF(\psi_1)$. Then Eq. (3) is
equivalent to the normalization
\be
\label{6}
N_0 \; = \; <\hat N_0> \; .
\ee
The statistical average of an operator $\hat A$ is defined in the standard
way as
$$
<\hat A> \; \equiv \; {\rm Tr}\; \hat\rho\; \hat A \; ,
$$
where $\hat\rho$ is a statistical operator and the trace is over $\cF(\psi_1)$.

One more restriction is
\be
\label{7}
<\psi_1(\br,t)> \; = \; 0 \; ,
\ee
which guarantees the conservation of quantum numbers. This can also be
rewritten as the quantum conservation condition
\be
\label{8}
<\hat\Lbd> \; = \; 0
\ee
for the self-adjoint operator
\be
\label{9}
\hat\Lbd \equiv \int \left [ \lbd(\br,t)\psi_1^\dgr(\br,t) +
\lbd^*(\br,t)\psi_1(\br,t) \right ]\; d\br \; ,
\ee
in which $\lbd(\br,t)$ is a complex function.

Two other common conditions is the normalization of the statistical operator
$\hat\rho$,
\be
\label{10}
<\hat 1_\cF> \; = \; 1 \; ,
\ee
and the definition of the internal energy
\be
\label{11}
E\; = \; <\hat H>
\ee
as the average of the Hamiltonian energy operator
$\hat H\equiv\hat H[\hat\psi]$, which is a functional of the shifted field
operator (1).

An equilibrium statistical ensemble for a Bose-condensed system is the pair
$\{\cF(\psi_1),\hat\rho\}$ of the space of microstates $\cF(\psi_1)$ and a
statistical operator $\hat\rho$. The notion of a representative ensemble stems
from the works of Gibbs [20], who emphasized that for the correct description
of the given statistical system, in addition to the standard conditions (10)
and (11), it is necessary to take into account all other constraints that
uniquely define the considered system. The corresponding statistical operator
can be found from the maximization of the Gibbs entropy $S\equiv -{\rm Tr}
\hat\rho\ln\hat\rho$ under the given statistical conditions. The conditional
maximization of the entropy is equivalent to the unconditional minimization
of the information functional [16,21]. In the present case, in addition to
conditions (10) and (11), we must also take into account the normalization
conditions (5) and (6) and the conservation constraint (8). Hence the
information functional is
$$
I[\hat\rho] = {\rm Tr}\; \hat\rho\; \ln\hat\rho + \lbd_0({\rm Tr}\;\hat\rho
-1 ) +
$$
\be
\label{12}
+ \bt ( {\rm Tr}\;\hat\rho \; \hat H -  E ) -
\bt\mu_0 ( {\rm Tr}\;\hat\rho \; \hat N_0 - N_0 ) -
\bt \mu_1 ( {\rm Tr}\;\hat\rho \; \hat N_1  - N_1 ) -
\bt {\rm Tr}\;\hat\rho\; \hat\Lbd \; ,
\ee
in which $\lbd_0$, $\bt$, $\bt\mu_0$, $\bt\mu_1$, and $\bt\lbd$ are the
appropriate Lagrange multipliers guaranteeng the validity of conditions (5),
(6), (8), (10), and (11). Minimizing functional (12), we get the statistical
operator
\be
\label{13}
\hat\rho = \frac{1}{Z}\; e^{-\bt H} \; ,
\ee
with the inverse temperature $\bt\equiv 1/T$, the partition function
$$
Z = \exp(1+\lbd_0) = {\rm Tr}\; e^{-\bt H} \; ,
$$
and the grand Hamiltonian
\be
\label{14}
H \equiv \hat H - \mu_0 \hat N_0 - \mu_1 \hat N_1 - \hat \Lbd \; .
\ee

Let us explain in more detail the role played by the Lagrange multipliers
$\mu_0$ and $\mu_1$. Above the Bose-Einstein condensation point, when
$N_0=0$ and $\hat N_1=\hat N$, one needs just one Lagrange multiplier
$\mu_1=\mu$, coinciding with the system chemical potential, whose role
is to preserve the total average number of particles $N=<\hat N>$. However,
below the condensation point, when the global gauge symmetry is broken,
there are two types of particles, condensed and uncondensed ones. The
number of condensed particles $N_0$, according to the Bogolubov theory
[8,9,17,18], has to be such that to make the system stable by minimizing
the thermodynamic potential. For an equilibrium statistical system with
the statistical operator (13), the grand thermodynamic potential is
\be
\label{15}
\Om = - T\ln {\rm Tr} e^{-\bt H} \; ,
\ee
with the grand Hamiltonian (14). Extremizing the grand potential (15) with
respect to the number of condensed particles, from the equation
$$
\frac{\prt\Om}{\prt N_0}= 0
$$
one obtains
$$
\mu_0= \; < \frac{\prt\hat H}{\prt N_0} >\; .
$$
This means that the Lagrange multiplier $\mu_0$ is responsible for the
thermodynamic stability of the system. Another Lagrange multiplier, $\mu_1$,
guarantees the normalization condition (5) for the number of uncondensed
particles $N_1$. But the latter, since $N_1=N-N_0$, implies that $\mu_1$
preserves the total average number of particles $N$. In this way, for a
Bose-condensed system, contrary to the system without the Bose-Einstein
condensate, there are two conditions on the number of particles. One
condition, as earlier, is that the total number of particles be $N$. And
another condition is that the number of condensed particles, $N_0$, would
be such that to provide the stability of the system by minimizing the
thermodynamic potential. This is why one needs two Lagrange multipliers
in order to guarantee the validity of these two conditions at each step of
any calculational procedure. As is shown below by practical calculations,
the use of two Lagrange multipliers makes the theory self-consistent,
avoiding the Hohenberg-Martin dilemma and yielding the results that are
in agreement with those derived analytically for the weak-coupling limit
as well as obtained by Monte-Carlo computer simulations for strong
interactions.

It is also important to keep in mind that the introduction of Lagrange
multipliers is a technical method allowing us to simplify calculations.
The number of the introduced multipliers is connected with the concrete
properties of the employed approach. Thus, in the Bogolubov theory
[8,9,17,18] one deals with two independent field variables, the condensate
wave function $\eta(\br,t)$ and the operator of uncondensed particles
$\psi_1(\br,t)$. This is why, as is explained above, it is convenient to
introduce two Lagrange multipliers.

One could ask whether we could limit ourselves by introducing a sole
Lagrange multiplier. The answer is straightforward: Yes, we could, but
the calculational procedure should then be changed. For instance, we could
follow the way of Hugenholtz and Pines [22], which was also used by
Gavoret and Nozieres [23]. They consider a uniform equilibrium system at
zero temperature, defining the number of condensed particles
$N_0=N_0(\rho,T)$ as a function of density $\rho$ and temperature $T$ from
the extremization of the internal energy $E\equiv<\hat H>$. The found
$N_0(\rho,T)$ is substituted explicitly into the Hamiltonian $\hat H$,
after which one works with the grand Hamiltonian $H=\hat H-\mu\hat N_1$,
where $\hat N_1$ is the operator for the number of  {\it uncondensed}
particles. Since the number of condensed atoms has already been defined
earlier from the stability condition, one requires now to use the sole
Lagrange multiplier aiming at guaranteeing the normalization condition
$N_1=<\hat N_1>$ for uncondensed particles, hence, because of the fixed
relation $N_1=N-N_0$, preserving the total number of particles $N$. This
way of calculations, with details expounded in Refs. [22,23], is
mathematically equivalent to the procedure, when $N_0$ has not been fixed
in advance but, instead, a Lagrange multiplier $\mu_0$ is introduced to
guarantee the stability condition in the process of calculations, after
which the number of condensed particles is defined as $N_0=N-N_1$, which
specifies the condensate depletion. It is this method of using an additional
Lagrange multiplier that is employed in the present paper.

Even more, it could be possible to work introducing no any Lagrange
multipliers at all, but which again would necessitate a different
calculational procedure. Thus, one could use the Girardeau-Arnowitt
approach [24,25] in the frame of the canonical ensemble, which requires
to invoke the so-called number-conserving field operators. The weak
point of this approach is that, as is well known [24,25], it yields the
unphysical gap in the spectrum, when approximate calculations are involved.
Girardeau mentioned [26] that an exact theory should not have the gap.
Later, Takano [27] demonstrated that, really, the gap should not arise when
all terms of the Hamiltonian are taken into account. Unfortunately, there
are no exact solutions for a realistic interacting Bose system, so that one
always has to resort to some approximations, in the course of which the
gap again reappears. This is why it is more convenient to work with the
grand canonical ensemble, introducing the Lagrange multipliers that would
assure the self-consistency of any calculational scheme.

Thus, the method of Lagrange multipliers has to be treated as a technical
procedure allowing us to simplify calculations. The number of the
required multipliers is intimately related to the concrete calculational
details. In the case of the Bogolubov approach [8,9,17,18], introducing
two different field variables, the condensate function $\eta$ and the
field operator of uncondensed particles $\psi_1$, it is convenient to
define two Lagrange multipliers that guarantee the validity of two
conditions, the thermodynamic stability condition and the conservation
of the total average number of particles.

When two Lagrange multipliers ate involved, none of them plays the role
of the system chemical potential. To define the latter, we may proceed as
follows. Keeping in mind that in experiments the total number of particles
is usually fixed, we may write for the internal energy (11) the standard
relation
\be
\label{16}
E = \; <H> + \;\mu N \; ,
\ee
connecting $E$ with the average of the grand Hamiltonian $<H>$ and with
the system chemical potential $\mu$. At the same time, substituting into
Eq. (11) the grand Hamiltonian (14), and taking into account condition
(8), we come to the expression
\be
\label{17}
E = \; <H> +\mu_0 N_0 + \mu_1 N_1 \; .
\ee

Comparing Eqs. (16) and (17) gives the definition of the system chemical
potential
\be
\label{18}
\mu \equiv \mu_0 n_0 + \mu_1 n_1
\ee
expressed through the Lagrange multipliers $\mu_0$ and $\mu_1$ and the
related atomic fractions
$$
n_0 \equiv \frac{N_0}{N} \; , \qquad n_1 \equiv \frac{N_1}{N} \; .
$$

The equations of motion for the variables $\eta$ and $\psi_1$ are given in
the usual manner as
\be
\label{19}
i\; \frac{\prt}{\prt t} \; \eta(\br,t) = \frac{\dlt H}{\dlt\eta^*(\br,t)}
\ee
for the condensate function and, respectively, as
\be
\label{20}
i\; \frac{\prt}{\prt t}\; \psi_1(\br,t) =
\frac{\dlt H}{\dlt\psi_1^\dgr(\br,t)}
\ee
for the field operator of uncondensed atoms.

In this way, the representative statistical ensemble for an arbitrary
equilibrium Bose system with broken gauge symmetry is the pair
$\{\cF(\psi_1),\hat\rho\}$ of the Fock space of microstates $\cF(\psi_1)$
and the statistical operator (13) with the grand Hamiltonian (14). The
notion of representative statistical ensembles can be extended to
nonequilibrium systems by considering the extremization of an effective
action functional [28].

\section{Bose Gas with Contact Interactions}

To specify the consideration, let us take the Hamiltonian energy operator in
the usual form
\be
\label{21}
\hat H = \int \hat\psi^\dgr(\br) \left ( -\;
\frac{\nabla^2}{2m}\right ) \hat\psi(\br)\; d\br +
\frac{1}{2} \; \Phi_0 \int \hat\psi^\dgr(\br) \hat\psi^\dgr(\br)
\hat\psi(\br)\hat\psi(\br)\; d\br \; ,
\ee
corresponding to the contact interaction potential with the strength
\be
\label{22}
\Phi_0 \equiv 4\pi\; \frac{a_s}{m} \; ,
\ee
where $a_s$ is the scattering length and $m$, mass. Here
$\hat\psi(\br)=\hat\psi(\br,t)$ is the shifted field operator (1).

The evolution equation for the condensate function is obtained by averaging
Eq. (19). To this end, we need the notation for the local condensate density
\be
\label{23}
\rho_0(\br,t) \equiv |\eta(\br,t)|^2 \; ,
\ee
normal density of uncondensed atoms
\be
\label{24}
\rho_1(\br,t) \; \equiv \; <\psi_1^\dgr(\br,t)\psi_1(\br,t)> \; ,
\ee
anomalous density
\be
\label{25}
\sgm_1(\br,t) \; \equiv \; <\psi_1(\br,t)\psi_1(\br,t)> \; ,
\ee
and the triple correlator
\be
\label{26}
\xi(\br,t) \; \equiv \;
<\psi_1^\dgr(\br,t)\psi_1(\br,t)\psi_1(\br,t)> \; .
\ee
The total local density is
\be
\label{27}
\rho(\br,t) = \rho_0(\br,t) + \rho_1(\br,t) \; .
\ee
Then, averaging Eq. (19) yields the evolution equation for the condensate
function
$$
i\; \frac{\prt}{\prt t}\; \eta(\br,t)  =\left ( -\;
\frac{\nabla^2}{2m} \; - \; \mu_0 \right ) \eta(\br,t) +
$$
\be
\label{28}
+ \Phi_0 \left [ \rho(\br,t) \eta(\br,t) + \rho_1(\br,t)\eta(\br,t) +
\sgm_1(\br,t)\eta^*(\br,t) + \xi(\br,t) \right ] \; .
\ee

The equation of motion for the field operator of uncondensed atoms follows
from Eq. (20) giving
$$
i\; \frac{\prt}{\prt t}\; \psi_1(\br,t) = \left ( -\; \frac{\nabla^2}{2m} \;
- \; \mu_1\right ) \psi_1(\br,t) +
$$
\be
\label{29}
+ \Phi_0 \left [ 2\rho_0(\br,t)\psi_1(\br,t) +
\eta^2(\br,t)\psi_1^\dgr(\br,t) + \hat X(\br,t) \right ] \; ,
\ee
where the last term is the correlation operator
$$
\hat X(\br,t) \equiv \left [ 2\psi_1^\dgr(\br,t) \eta(\br,t) +
\eta^*(\br,t)\psi_1(\br,t) + \psi_1^\dgr(\br,t)\psi_1(\br,t)
\right ]\psi_1(\br,t) \; .
$$

In equilibrium state, one has
\be
\label{30}
\frac{\prt}{\prt t}\; \eta(\br,t) = 0 \; .
\ee
Also, if the system is uniform, then
$$
|\eta(\br,t)|^2 = \frac{N_0}{V} \equiv \rho_0 \; ,
\qquad \rho_1(\br,t) = \frac{N_1}{V}  =\rho_1 \; ,
$$
\be
\label{31}
\sgm_1(\br,t)\equiv \sgm_1 \; , \qquad \xi(\br,t)\equiv\xi \; ,
\qquad \rho \equiv \frac{N}{V} = \rho_0 + \rho_1 \; .
\ee
In that case, equation (28) for the condensate function gives the Lagrange
multiplier
\be
\label{32}
\mu_0 =\left ( \rho + \rho_1 +\sgm_1 +
\frac{\xi}{\sqrt{\rho_0}} \right ) \Phi_0 \; .
\ee

The field operator of uncondensed atoms can be expanded over a complete
basis. In general, if the system were nonuniform, in the presence of an
external potential, it would be convenient to take a basis formed by natural
orbitals [29,30]. For the uniform system under consideration, the natural
orbitals are just plane waves $\vp_k(\br)=e^{i\bk\cdot\br}/\sqrt{V}$. The
corresponding expansion of $\psi_1$ reads as
$$
\psi_1(\br,t) = \sum_{k\neq 0} a_k(t) \vp_k(\br) \; .
$$
With this expansion, the grand Hamiltonian (14) takes the form of a sum
\be
\label{33}
H = \sum_{n=0}^4 H^{(n)}
\ee
of five terms, classified according to the number of the operators $a_k$
or $a_k^\dgr$ in the products. The zero-order term does not contain $a_k$,
\be
\label{34}
H^{(0)} = \left ( \frac{1}{2}\; \rho_0 \Phi_0 - \mu_0 \right ) N_0 \; .
\ee

Generally, in order to satisfy condition (8), it is necessary and
sufficient [31] that the Hamiltonian (14) would not contain the terms
linear in $\psi_1$ or $a_k$. This can be achieved by choosing the
corresponding Lagrange multipliers $\lbd(\br,t)$ in Eq. (9). For a
uniform system, because of the orthogonality condition (2), one has
$\hat\Lbd=0$ and $H^{(1)}=0$ automatically.

The second-order term is
\be
\label{35}
H^{(2)} = \sum_{k\neq 0} \left [ \left ( \frac{k^2}{2m} +
2\rho_0\Phi_0 - \mu_1\right ) a_k^\dgr a_k + \frac{1}{2}\;
\rho_0 \Phi_0 \left ( a_k^\dgr a_{-k}^\dgr + a_{-k}a_k
\right ) \right ] \; .
\ee
In the third-order term
\be
\label{36}
H^{(3)} = \sqrt{\frac{\rho_0}{V}} \; \Phi_0 \;
{\sum_{p,q}}' \left ( a_q^\dgr a_{q-p} a_p +
a_p^\dgr a_{q-p}^\dgr a_q \right ) \; ,
\ee
the prime on the summation symbol implies that ${\bf p}\neq 0$,
${\bf q}\neq 0$, and ${\bf p}-{\bf q}\neq 0$. The fourth-order term is
\be
\label{37}
H^{(4)} = \frac{\Phi_0}{2V} \; \sum_k {\sum_{p,q}}'
a_p^\dgr a_q^\dgr a_{k+p} a_{q-k} \; ,
\ee
where the prime means that ${\bf p}\neq 0$, ${\bf q}\neq 0$,
$\bk+{\bf p}\neq 0$, and $\bk-{\bf q}\neq 0$.

To proceed further, we need to invoke some approximation. A natural
mean-field approximation, in the presence of broken gauge symmetry, is the
HFB approximation. This is used to simplify the higher-order terms (36) and
(37). Employing the designations
\be
\label{38}
\om_k \equiv \frac{k^2}{2m} + 2\rho \Phi_0 - \mu_1
\ee
and
\be
\label{39}
\Dlt \equiv (\rho_0 +\sgm_1) \Phi_0 \; ,
\ee
we obtain
\be
\label{40}
H = E_{HFB} + \sum_{k\neq 0} \left [ \om_k a_k^\dgr a_k +
\frac{\Dlt}{2}\left ( a_k^\dgr a_{-k}^\dgr + a_{-k} a_k \right )
\right ] \; ,
\ee
where the nonoperator term is
\be
\label{41}
E_{HFB} =  H^{(0)} \; - \; \frac{\Phi_0}{2\rho}
\left ( 2\rho_1^2 + \sgm_1^2 \right ) N \; .
\ee

Hamiltonian (40) can be diagonalized by means of the Bogolubov canonical
transformation $a_k=u_kb_k+v^*_{-k}b^\dgr_{-k}$, in which
$$
u_k^2 = \frac{\om_k+\ep_k}{2\ep_k} \; , \qquad
v_k^2 = \frac{\om_k-\ep_k}{2\ep_k} \; ,
$$
with the Bogolubov spectrum
\be
\label{42}
\ep_k =\sqrt{\om_k^2 -\Dlt^2} \; .
\ee
Then one gets
\be
\label{43}
H = E_B + \sum_{k\neq 0} \ep_k b_k^\dgr b_k \; ,
\ee
where
\be
\label{44}
E_B \equiv E_{HFB} + \frac{1}{2}\; \sum_{k\neq 0} (\ep_k -\om_k)\; .
\ee

By the Bogolubov [18] and Hugenholtz-Pines [22] theorems, the spectrum is
to be gapless, which implies that
\be
\label{45}
\lim_{k\ra 0} \ep_k = 0 \; , \qquad \ep_k\geq 0\; .
\ee
From here it follows that
\be
\label{46}
\mu_1 = (\rho +\rho_1 -\sgm_1) \Phi_0 \; .
\ee
The condensate multiplier (32) in the HFB approximation, when $\xi=0$,
becomes
\be
\label{47}
\mu_0 = (\rho + \rho_1 +\sgm_1) \Phi_0 \; .
\ee
It is important to emphasize that the form of $\mu_1$ in Eq. (46) is
necessary and sufficient for making the spectrum gapless. That it is
sufficient follows at once after substituting Eq. (46) into the Bogolubov
spectrum (42). And the necessity stems from the Bogolubov theorem [18]
which can be formulated as the inequality
$$
\vert \mu_1 - \frac{k^2}{2m} + \Sigma_{12}(\bk,0) -
\Sigma_{11}(\bk,0) \vert \leq \frac{k^2}{2m_0} \; ,
$$
where $\Sigma_{12}$ and $\Sigma_{11}$ are the normal and anomalous
self-energies. Setting here $\bk=0$ leads to the Hugenholtz-Pines relation
$$
\mu_1 = \Sigma_{11}(0,0) - \Sigma_{12}(0,0) \; .
$$
In the HFB approximation, one has
$$
\Sigma_{11}(0,0) = 2\rho \Phi_0\; , \qquad \Sigma_{12}(0,0) =
(\rho_0 + \sgm_1) \Phi_0\; ,
$$
from where one immediately obtains Eq. (46).

With multiplier (46), spectrum (42) takes the form
\be
\label{48}
\ep_k = \sqrt{(ck)^2+\left ( \frac{k^2}{2m}\right )^2 } \; ,
\ee
in which the sound velocity is
\be
\label{49}
c \equiv \sqrt{\frac{\Dlt}{m}} \; .
\ee

For the diagonal Hamiltonian (43), it is straightforward to calculate all
averages, such as the normal average
\be
\label{50}
n_k \; \equiv \; <a_k^\dgr a_k>
\ee
and the anomalous average
\be
\label{51}
\sgm_k \; \equiv \; < a_k a_{-k}> \; .
\ee
Their integration over momenta gives the density of uncondensed particles
\be
\label{52}
\rho_1 = \int n_k \; \frac{d\bk}{(2\pi)^3}
\ee
and, respectively, the anomalous average
\be
\label{53}
\sgm_1 = \int \sgm_k \; \frac{d\bk}{(2\pi)^3} \; .
\ee
The quantity $|\sgm_1|$ can be interpreted as the density of pair-correlated
atoms [19].

In what follows, we concentrate on the zero-temperature properties of the
system. When $T=0$, then $<b_k^\dgr b_k>=0$. The normal average (50) is
\be
\label{54}
n_k = \frac{\om_k-\ep_k}{2\ep_k} \; ,
\ee
while the anomalous average (51) becomes
\be
\label{55}
\sgm_k = -\; \frac{\Dlt}{2\ep_k} \; .
\ee
Combining Eqs. (38), (39), (46), and (49), we have
$$
\om_k = \frac{k^2}{2m} + \Dlt \; , \qquad \Dlt =mc^2 \; .
$$
The equation for the sound velocity (49) can be represented as
\be
\label{56}
mc^2 = ( \rho_0 + \sgm_1)\Phi_0 \; .
\ee
Note that from here, in the limit of asymptotically weak interactions, we
get the Bogolubov expression
$$
c \simeq \sqrt{ \frac{\rho_0\Phi_0}{m}} \qquad (\Phi_0\ra 0) \; .
$$

For the density of uncondensed atoms (52), we find
\be
\label{57}
\rho_1 = \int \frac{\om_k-\ep_k}{2\ep_k} \; \frac{d\bk}{(2\pi)^3} =
\frac{(mc)^3}{3\pi^2} \; .
\ee
And for the anomalous average (53), we have
\be
\label{58}
\sgm_1 = -\; \frac{\Dlt}{2}\; \int \frac{1}{\ep_k}\;
\frac{d\bk}{(2\pi)^3} \; .
\ee

The integral $\int d\bk/\ep_k$ in Eq. (58) is ultraviolet divergent.
To overcome this, we can resort to the standard procedure of analytic
regularization [5,32]. For this purpose, we, first, consider the integral
in the limit of asymptotically small $\Phi_0$, when the dimensional
regularization is applicable, and then analytically continue the result
to arbitrary interactions. The dimensional regularization gives
$$
\int \frac{1}{\ep_k}\; \frac{d\bk}{(2\pi)^3} = -\;
\frac{2}{\pi^2}\; m^{3/2} \; \sqrt{\rho_0\Phi_0} \; .
$$
Using this in Eq. (58), we obtain
\be
\label{59}
\sgm_1 = \frac{(mc)^2}{\pi^2}\; \sqrt{m\rho_0\Phi_0} \; .
\ee

It is important to stress that the anomalous density (59) enjoys the natural
limiting property
\be
\label{60}
\sgm_1 \ra 0 \qquad (\rho_0\ra 0) \; .
\ee
The physics of property (60) is evident. The existence of both the condensate
density $\rho_0$ and the anomalous density $\sgm_1$ is caused by the gauge
symmetry breaking. Both of them are nonzero as soon as the symmetry is broken,
while both become zero if the symmetry is restored. Any of these quantities
could be treated as an order parameter for the broken-symmetry phase. So, both
these quantities, $\rho_0$ and $\sgm_1$, have to nullify simultaneously, when
one of them tends to zero.

We may also note that simplifying $\sgm_1$ by replacing $\sqrt{m\rho_0\Phi_0}$
by $mc$, as is done in Ref. [14], is admissible only in the limit of weak
interactions, when $\rho_0\sim\rho$, But for strong interactions, when
$\rho_0\ra 0$, this replacement does not hold, since then the limiting
property (60) is not satisfied. Therefore, the results of Ref. [14] are
quantitatively correct in the limit of weak interactions, though for strong
interactions, they may give only a qualitative picture. And our aim in the
present paper is to give a careful analysis of the system properties for
arbitrary strong interactions in the whole range of $\Phi_0\in[0,\infty)$.
This requires to employ Eq. (59), which explicitly satisfy the limiting
condition (60).

\section{System Characteristics at Varying Interactions}

The effective interaction strength can be characterized by the dimensionless
{\it gas parameter}
\be
\label{61}
\gm \equiv \rho^{1/3} a_s \; .
\ee
One often uses the quantity $\rho a_s^3$ as a parameter quantifying
the interaction strength. However, parameter (61), to our mind, is
more convenient, since it better distinguishes between weak and
strong interactions. Thus, the majority of experiments with ultracold
trapped gases [1--7] deals with weakly interacting atoms, so that
$\rho a_s^3\sim 10^{-8}-10^{-4}$, which corresponds to
$\gm\sim 10^{-3}-10^{-2}$. The effective interactions can be
noticeably strengthened by loading atoms in optical lattices [33].
Contrary to weakly interacting gases, superfluid $^4$He is a strongly
interacting system. The $^4$He atoms can be represented by hard spheres
of diameter $a_s$ [34,35]. At saturated vapour pressure, one has [35]
$a_s=2.139\AA$ and $\rho\approx 0.022\AA^{-3}$, hence $\rho a_s^3\approx
0.215$, which is yet much less than one. But the corresponding $\gm\approx
0.599$ is close to one. So, weak  interactions are characterized by
$\gm\ll 1$, while a strongly interacting system has $\gm\sim 1$.

It is convenient to introduce the dimensionless sound velocity
\be
\label{62}
s \equiv \frac{mc}{\rho^{1/3}} \; ,
\ee
the fraction of uncondensed atoms
\be
\label{63}
n_1 \equiv \frac{\rho_1}{\rho} = \frac{N_1}{N} \; ,
\ee
and the anomalous fraction
\be
\label{64}
\sgm \equiv \frac{\sgm_1}{\rho} \; .
\ee
In these dimensionless quantities, Eq. (56) writes as
\be
\label{65}
s^2 = 4\pi(n_0+\sgm)\gm  \; .
\ee
Here
\be
\label{66}
n_0 = 1-n_1
\ee
and
\be
\label{67}
n_1 = \frac{s^3}{3\pi^2} \; ,
\ee
while the anomalous fraction (64) is
\be
\label{68}
\sgm = \frac{2s^2}{\pi^{3/2}}\; \sqrt{\gm n_0 } \; .
\ee
Four quantities, $s$, $n_0$, $n_1$, and $\sgm$, are defined by the system
of four equations (65) to (68).

In the limit of asymptotically weak interactions, when $\gm\ra 0$ and
$\sgm\ra 0$, the condensate fraction and sound velocity tend to the
Bogolubov expressions
\be
\label{69}
n_B =  1 \; - \; \frac{8}{3\sqrt{\pi}}\; \gm^{3/2}
\ee
and, respectively,
\be
\label{70}
s_B = 2\sqrt{\pi\gm} \; .
\ee
For the higher orders with respect to $\gm$, we find
\be
\label{71}
n_0 \simeq 1 \; - \; \frac{8}{3\sqrt{\pi}}\; \gm^{3/2} \; - \;
\frac{64}{3\pi}\; \gm^3 \; - \; \frac{640}{9\pi^{3/2}}\; \gm^{9/2} \; ,
\ee
\be
\label{72}
s \simeq 2\sqrt{\pi}\; \gm^{1/2} + \frac{16}{3}\; \gm^2 +
\frac{32}{9\sqrt{\pi}}\; \gm^{7/2} \; - \; \frac{3904}{27\pi}\; \gm^5 \; ,
\ee
\be
\label{73}
\sgm \simeq \frac{8}{\sqrt{\pi}} \; \gm^{3/2} + \frac{32}{\pi}\;\gm^3 \; -
\; \frac{64}{\pi^{3/2}}\; \gm^{9/2} \; .
\ee

For asymptotically strong interactions, when $\gm\ra\infty$, we obtain
\be
\label{74}
n_0 \simeq \frac{\pi}{64}\; \gm^{-3} \; - \; \frac{1}{512} \left (
\frac{\pi^5}{9}\right )^{1/3} \gm^{-5} \; ,
\ee
\be
\label{75}
s \simeq (3\pi^2)^{1/3}\; - \; \frac{1}{64} \left (
\frac{\pi^5}{9}\right )^{1/3} \gm^{-3} + \frac{1}{1536} \left (
\frac{\pi^7}{3}\right )^{1/3} \gm^{-5} \; ,
\ee
\be
\label{76}
\sgm \simeq \frac{(9\pi)^{1/3}}{4} \; \gm^{-1} \; - \;
\frac{\pi}{64}\;\gm^{-3} \; - \; \frac{1}{128}\left (
\frac{\pi^4}{3}\right )^{1/3} \gm^{-4} + \frac{1}{512} \left (
\frac{\pi^5}{9} \right )^{1/3} \gm^{-5} \; .
\ee
So, the condensate fraction tends to zero, together with the anomalous
fraction, as $\gm\ra\infty$, in agreement with condition (60).

For the whole region of $\gm$, we solve numerically the system of Eqs. (65)
to (68), showing the results in Figs. 1,2, and 3.

Figure 1 presents the condensate fraction (66) and this fraction (69) in the
Bogolubov approximation. At small $\gm$, up to $\gm\approx 0.1$, $n_0$ and
$n_B$ practically coincide with each other. For $\gm >0.1$, the Bogolubov
approximation overestimates the condensate fraction. But $n_B=0$ at
$\gm=0.762$, while $n_0$ is yet finite, though small. The condensate
fraction of a homogeneous Bose gas, at zero temperature, as a function of
the gas parameter was calculated by means of the Monte Carlo simulation by
Giorgini et al. [36] up to $\gm\approx 0.5$. For this region of
$\gm$, our $n_0$ is in good agreement with the Monte Carlo calculations.
Monte Carlo techniques have also been used for studying the condensate
fraction of trapped atoms [37--39]. But the latter results cannot be
directly compared with $n_0$ in a homogeneous gas, since in traps, the
condensate fraction is a function of spatial variables as well as of the
trap shape [37--41]. What is possible and interesting to compare is the
condensate fraction in superfluid $^4$He at zero temperature and our
$n_0$ at $\gm\approx 0.6$ corresponding to liquid $^4$He with hard-core
interactions. For $\gm\approx 0.6$, we have $n_0\approx 0.15$, which is
close to the condensate fraction $n_0\approx 0.1$, measured in experiments
(as is reviewed in Refs. [3,42]) as well as obtained by Monte Carlo
simulations (see review articles [43,44]).

Figure 2 shows the dimensionless sound velocity (62) and its Bogolubov
approximation (70). These quantities practically coincide up to $\gm\approx
0.1$. For $\gm>0.1$ the Bogolubov form $s_B$ underestimates $s$ till
$\gm\approx 0.7$, after which it overestimates the latter.

In Fig. 3, we compare the fraction of uncondensed atoms (67) with the
anomalous fraction (68). As is seen, $\sgm>n_1$ up to $\gm\approx 0.7$. The
anomalous fraction $\sgm$ becomes substantially smaller than $n_1$ only for
very large $\gm\gg 1$. This is in agreement with other calculations [45]
confirming that anomalous averages cannot be neglected at low temperatures.

Let us now analyse the ground-state energy
\be
\label{77}
E\; \equiv \; <H> + \mu N \qquad (T=0) \; .
\ee
According to Eq. (43),
$$
<H>\; = \; E_B \qquad (T=0) \; .
$$
The system chemical potential, defined in Eq. (18), is expressed through the
Lagrange multipliers (46) and (47) and the fractions $n_0$ and $n_1$, which
gives
\be
\label{78}
\mu = (1+n_1+\sgm-2\sgm n_1)\rho\Phi_0 \; .
\ee
For the dimensionless chemical potential
\be
\label{79}
\overline\mu \equiv \frac{2m\mu}{\rho^{2/3}} \; ,
\ee
we get
\be
\label{80}
\overline\mu = 8\pi\gm(1+n_1+\sgm-2\sgm n_1) \; .
\ee
From Eq. (44), we have
\be
\label{81}
E_B = E_{HFB} + N \int \frac{\ep_k-\om_k}{2\rho}\;
\frac{d\bk}{(2\pi)^3} \; ,
\ee
where the integral is calculated invoking the dimensional regularization
[5], giving
$$
\int \frac{\ep_k-\om_k}{2\rho}\; \frac{d\bk}{(2\pi)^3} =
\frac{8(mc)^5}{15\pi^2m\rho} \; .
$$
Equations (34) and (41) yield
\be
\label{82}
\frac{E_{HFB}}{N} = \frac{\rho\Phi_0}{2}\left ( n_0^2 - 2n_1^2 -\sgm^2
\right ) - \mu_0 n_0 \; .
\ee
Summarizing these formulas, we find
\be
\label{83}
\frac{E}{N} = \frac{\rho\Phi_0}{2}\left ( 1 + n_1^2 - 2\sgm n_1 -
\sgm^2 \right ) + \frac{8(mc)^5}{15\pi^2m\rho} \; .
\ee
It is convenient to define the dimensionless ground-state energy
\be
\label{84}
E_0 \equiv \frac{2mE}{\rho^{2/3}N} \; ,
\ee
for which we obtain
\be
\label{85}
E_0 =  4\pi\gm \left ( 1 +n_1^2 - 2\sgm n_1 -\sgm^2 +
\frac{4s^5}{15\pi^3\gm}  \right ) \; .
\ee
This can be compared with the Lee-Huang-Yang approximation [46--48]
\be
\label{86}
E_{LHY} = 4\pi\gm \left ( 1 +
\frac{128}{15\sqrt{\pi}}\; \gm^{3/2} \right ) \; ,
\ee
derived for small $\gm$.

In the limits of asymptotically small and large $\gm$, the dimensionless
chemical potential (80) tends to
\be
\label{87}
\overline\mu \ra 8\pi\gm + \frac{256}{3}\; \sqrt{\pi}\; \gm^{5/2}
\qquad (\gm\ra 0)
\ee
and, respectively,
\be
\label{88}
\overline\mu \ra 16\pi\gm \qquad (\gm\ra\infty) \; .
\ee
The dimensionless ground-state energy (85) at small $\gm\ll 1$ possesses the
expansion
\be
\label{89}
E_0 \simeq 4\pi\gm + \frac{512}{15}\; \sqrt{\pi}\; \gm^{5/2} +
\frac{512}{9}\; \gm^4 \; ,
\ee
which reproduces the Lee-Huang-Yang approximation (86) for $\gm\ra 0$. And
for large $\gm\gg 1$, we find
\be
\label{90}
E_0 \simeq 8\pi\gm + \frac{6}{5}\left ( 9\pi^4\right )^{1/3} \; - \;
\frac{3}{4}\left ( 3\pi^5\right )^{1/3}\gm^{-1} +
\frac{1}{64}\left ( 3\pi^8\right )^{1/3}\gm^{-4} \; .
\ee
Note that expressions (87) and (88) can also be obtained from Eq. (89)
and (90) using the relation $\mu=\prt E/\prt N$, valid for $T=0$.

Figure 4 illustrates the behaviour of the ground-state energy (85) and
the Lee-Huang-Yang approximation (86). The latter is known to practically
coincide with the energy calculated trough the Monte Carlo simulations
[36] up to $\gm\approx 0.4$. As is seen from the figure, our $E_0$ is also
very close to $E_{LHY}$ in the region $0\leq\gm < 0.4$, but is lower than
$E_{LHY}$ for $\gm>0.4$. Hence, $E_0$ well reproduces the available data
of Monte Carlo calculations up to $\gm\approx 0.4$.

\section{Conclusion}

The notion of representative statistical ensembles is applied to Bose
systems with broken global gauge symmetry. A general procedure is described
for constructing the grand Hamiltonian for the representative ensemble of
an arbitrary equilibrium Bose system. A self-consistent mean-field theory is
developed, which is both conserving and gapless. The properties of a uniform
Bose gas at zero temperature are studied both analytically and numerically
for the gas parameter varying between zero and infinity. Thus, in the frame
of the suggested approach, strongly interacting systems can also be considered.
For instance, as is known [35,43], some of the properties of superfluid
$^4$He can be understood by treating the potential as a hard-core interaction
of diameter $a_s=2.139\AA$, which, at saturated vapor pressure, corresponds
to $\gm\approx 0.6$. For the latter $\gm$, we find the condensate fraction
$n_0$ of order $10\%$, which agrees with the condensate fraction in helium
at zero temperature, measured in experiments [42] and found in Monte Carlo
simulations [43], being also of order $10\%$. Another application of the
developed self-consistent mean-field theory with arbitrary strong
interactions could be the description of Bose-Einstein condensation of
multiquark clusters in nuclear matter [49,50].

\vskip 5mm

{\bf Acknowledgement}

\vskip 3mm

One of the authors (V.I.Y.) is grateful for financial support to the
German Research Foundation and for discussions to M. Girardeau, R. Graham,
and H. Kleinert.

\newpage

{\Large{\bf Figure Captions}}

\vskip 1cm

{\bf Fig. 1}. Condensate fraction $n_0$ (solid line) and its Bogolubov
approximation $n_B$ (dashed line) as functions of the gas parameter $\gm$.

\vskip 5mm

{\bf Fig. 2}. Dimensionless sound velocity $s$ (solid line) and its
Bogolubov approximation $s_B$ (dashed line) as functions of the gas
parameter $\gm$.

\vskip 5mm

{\bf Fig. 3}. Fraction of uncondensed atoms $n_1$ (dashed line) and
anomalous fraction $\sgm$ (solid line) as functions of the gas parameter
$\gm$.

\vskip 5mm

{\bf Fig. 4}. Dimensionless ground-state energy $E_0$ (solid line) and the
Lee-Huang-Yang approximation $E_{LHY}$ (dashed line) as functions of the gas
parameter $\gm$.

\newpage

\begin{figure}[h]
\centerline{\epsfig{file=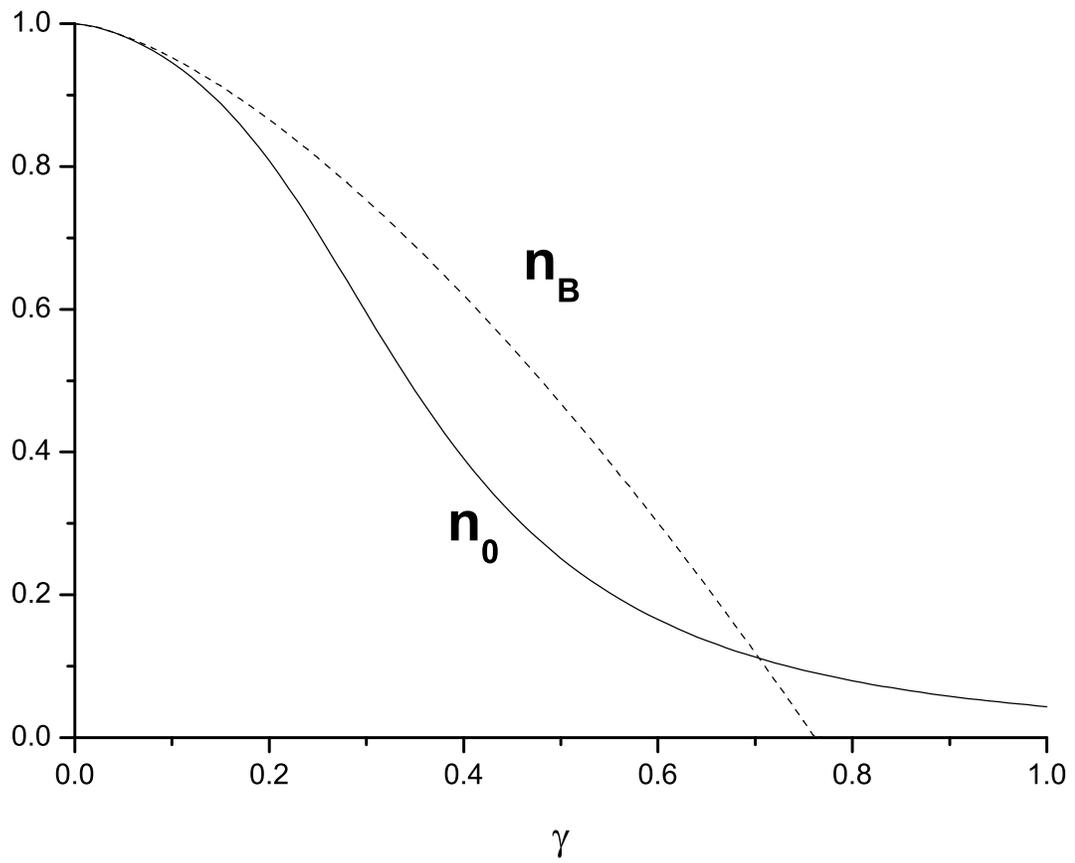,angle=0,width=16cm}}
\vskip 1cm
\caption{Condensate fraction $n_0$ (solid line) and its Bogolubov
approximation $n_B$ (dashed line) as functions of the gas parameter $\gm$.
}
\label{fig:Fig.1}
\end{figure}

\newpage

\begin{figure}[h]
\centerline{\epsfig{file=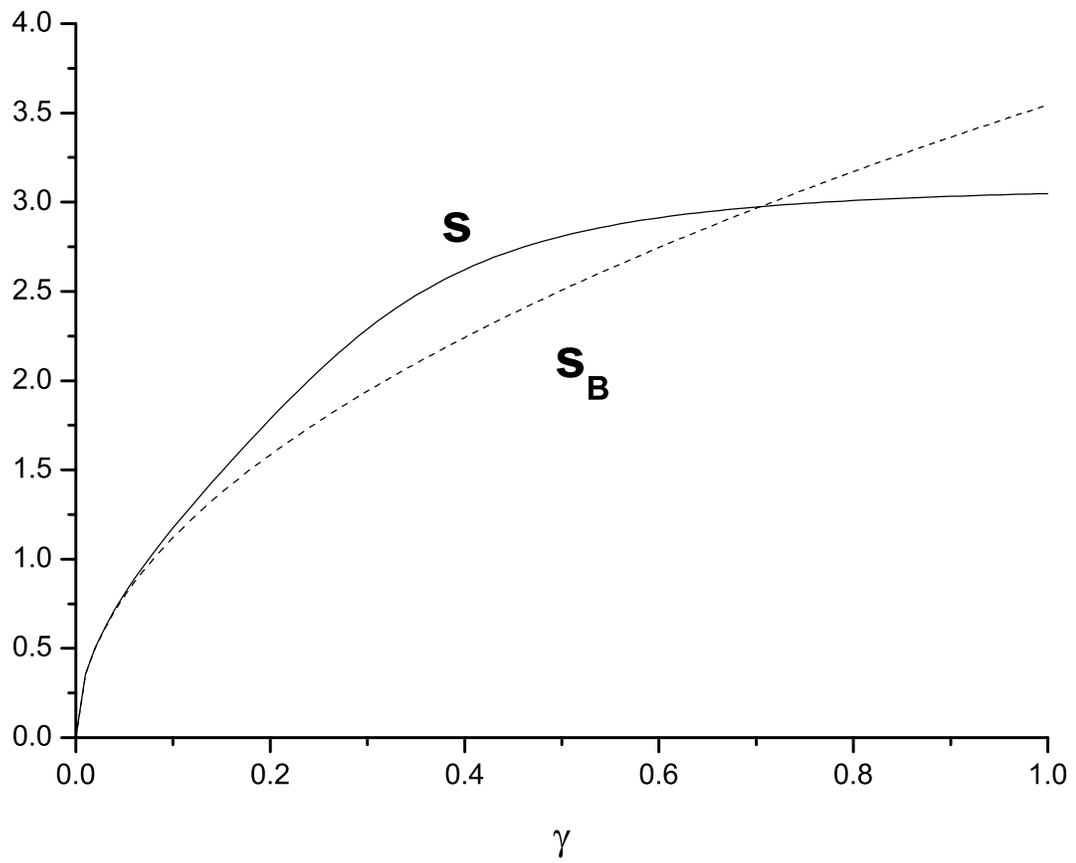,angle=0,width=16cm}}
\vskip 1cm
\caption{Dimensionless sound velocity $s$ (solid line) and its
Bogolubov approximation $s_B$ (dashed line) as functions of the gas
parameter $\gm$.
}
\label{fig:Fig.2}
\end{figure}

\newpage

\begin{figure}[h]
\centerline{\epsfig{file=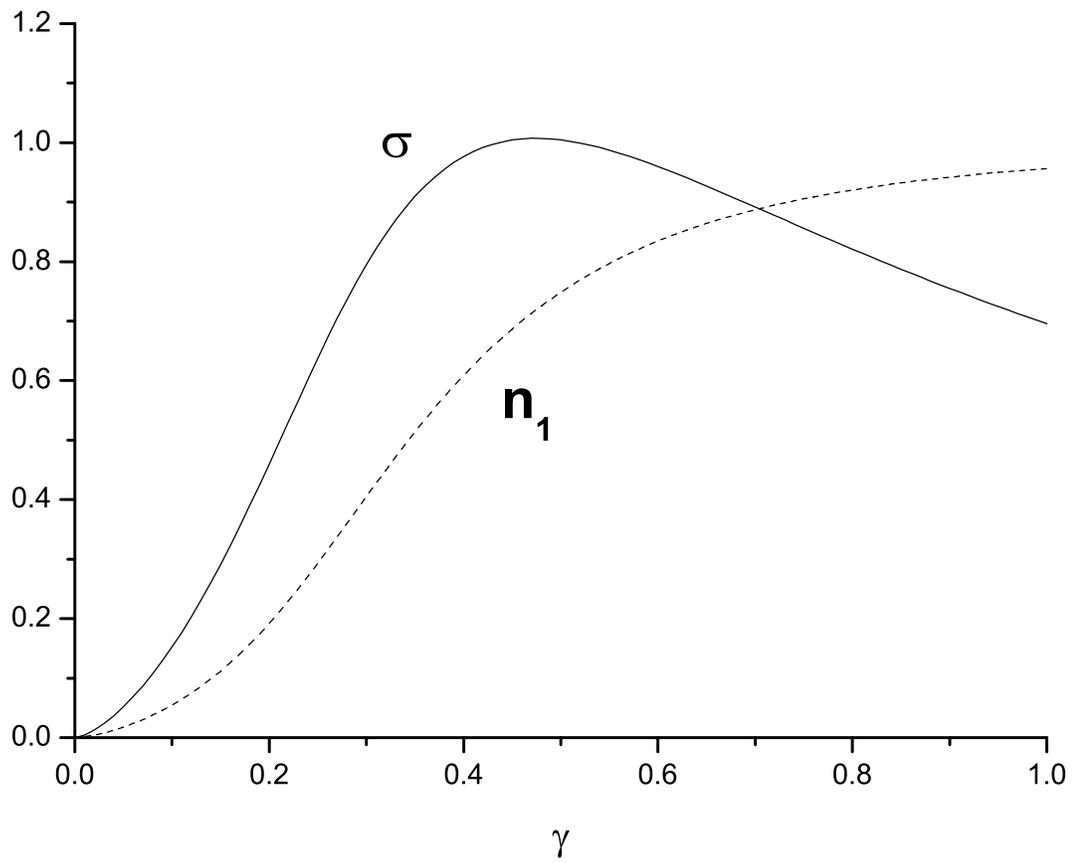,angle=0,width=16cm}}
\vskip 1cm
\caption{Fraction of uncondensed atoms $n_1$ (dashed line) and
anomalous fraction $\sgm$ (solid line) as functions of the gas parameter
$\gm$.
}
\label{fig:Fig.3}
\end{figure}

\newpage

\begin{figure}[h]
\centerline{\epsfig{file=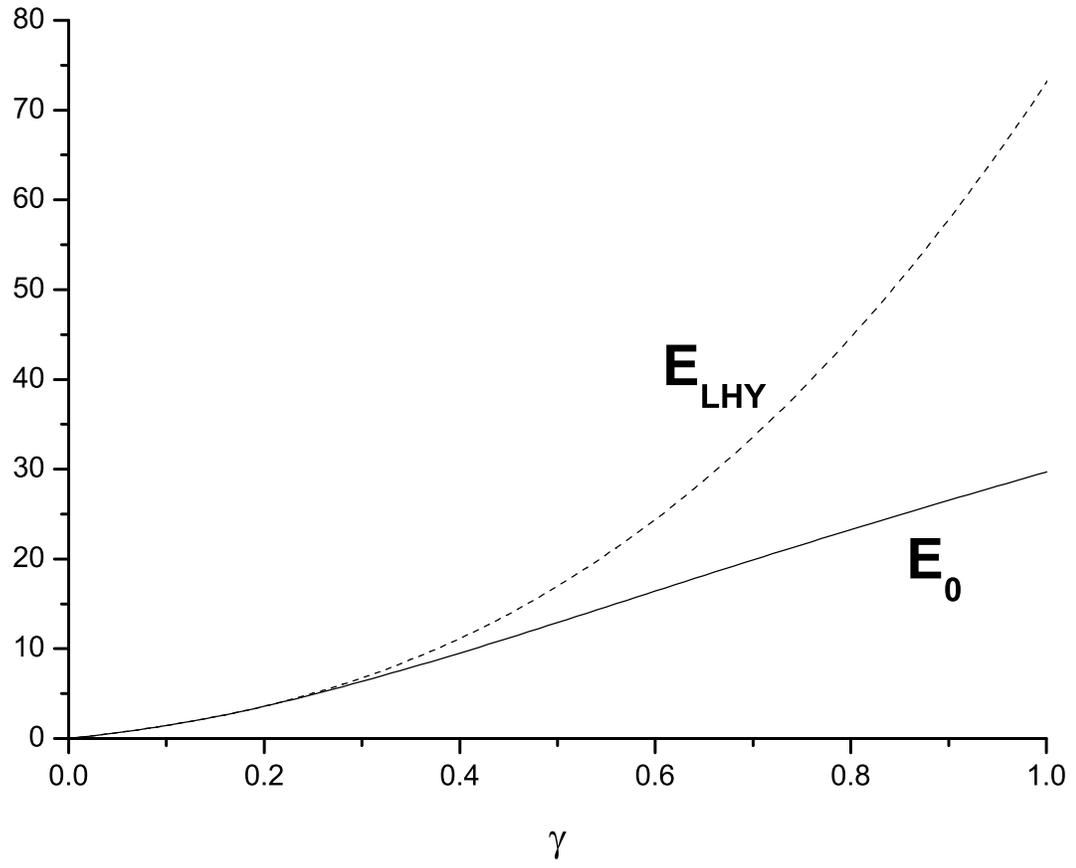,angle=0,width=16cm}}
\vskip 1cm
\caption{Dimensionless ground-state energy $E_0$ (solid line) and the
Lee-Huang-Yang approximation $E_{LHY}$ (dashed line) as functions of the gas
parameter $\gm$.
}
\label{fig:Fig.4}
\end{figure}

\end{document}